\documentstyle[12pt,epsf,psfig]{article}
\input epsf
\def\beq{\begin{equation}}
\def\eeq{\end{equation}}
\def\eeqn{\end{equation}}
\def\beqa{\begin{eqnarray}}
\def\eeqa#1{\label{#1}\end{eqnarray}}
\def\eeqan{\end{eqnarray}}

\def\np{\nu_k^{(+)}}
\def\nm{\nu_k^{(-)}}
\def\vp{\varphi}

\begin{document}

\vspace{0.1in}
\begin{center}
\bigskip\bigskip
{\large \bf Solitonic $D$-branes and brane annihilation}

\vspace{0.5in}      


\vskip 1cm {Gia Dvali$^a$ and Alexander Vilenkin$^{b,}$}

\vskip 1cm
{\it $^a$Department of Physics, Center for Cosmology and Particle Physics, New York University, New York, NY 10003\\
$^b$ Institute of Cosmology, Department of Physics and Astronomy,
Tufts University, Medford, MA 02155, USA}\\
\end{center}

\vspace{0.9cm}
\begin{center}
{\bf Abstract}
\end{center}

We point out some intriguing analogies between field theoretic
solitons (topological defects) and $D$-branes.  Annihilating
soliton-antisoliton pairs can produce stable solitons of lower
dimensionality.  Solitons that localize massless gauge fields in their
world volume automatically imply the existence of open flux tubes
ending on them and closed flux tubes propagating in the bulk.  We
discuss some aspects of this localization on explicit examples of
unstable wall-anti-wall systems.  The annihilation of these walls can
be described in terms of tachyon condensation which renders the
world-volume gauge field non-dynamical. During this condensation the
world volume gauge fields (open string states) are {\it resonantly}
excited.  These can later decay into closed strings, or get squeezed
into a network flux tubes similar to a network of cosmic strings
formed at a cosmological phase transition.  Although, as in the
$D$-brane case, perturbatively one can find exact time-dependent
solutions, when the energy of the system stays localized in the plane
of the original soliton, such solutions are unstable with respect to
decay into open and closed string states.  Thus, when a pair of such
walls annihilates, the energy is carried away (at least) by closed
string excitations (``glueballs''), which are the lowest energy
excitations about the bulk vacuum.  Suggested analogies can be useful
for the understanding of the complicated $D$-brane dynamics and of the
production of topological defects and reheating during brane collision
in the early universe.
     
\vspace{0.1cm}

\vspace{0.1in}

\section{Branes and Solitons}

Unstable $D$-brane systems may play an important role in cosmology,
since they can give an explicit string realization of the inflationary
scenario \cite{dvalitye} .  Their post -inflationary evolution can
provide simple mechanisms for baryogenesis \cite{giga}, and for
generation of dark matter in the form of open strings stretched
between the $D$-branes \cite{darkmatterdvali}.  Significant progress
has been recently made in the understanding of the dynamics of
unstable non-BPS $D$-brane configurations \cite{tachyon}.  Some
interesting features have emerged \cite{darkmatter}, which very
briefly can be summarized as follows.
 
Annihilation of $D$-brane anti-$D$-brane pair can be described in the
language of a (single) tachyon condensation. This tachyon is the
tachyon of the bosonic open string theory, living on the world-volume
of an unstable non-BPS brane. Tachyonic vacuum is the closed string
vacuum, about which there are no open string excitations. Due to this,
the energy initially stored in $D$-anti-$D$ system cannot dissipate in
the bulk, and stays localized in the plane of the original
$D$-brane. Energy-momentum of a rolling tachyon asymptotically reach
the form characteristic of a pressureless gas.

When the tachyon rolls away from the  maximum of its potential,
it can produce topologically stable knots in the world volume. These
knots are stable BPS $D$-branes of lower dimensionality. Thus,
annihilating $D$-branes can produce branes of lower dimensions.

The above features can have important cosmological implications in the
context of $D$-brane inflation \cite{dvalitye} (see
\cite{models,henry,braneangles} for recent progress in model
building). In this setup, a pair (or number) of originally separated
$D$-branes or brane-anti-branes slowly fall towards one another.
During this process the Universe undergoes a stage of exponential
expansion.  The scalar excitation corresponding to the inter-brane
distance plays the role of a slowly rolling inflaton field. Gradually
$D$-branes accelerate, and slow-roll inflation stops. Soon after this
point, branes collide and reheat the Universe. This marks the beginning
of the standard big bang cosmology, in which the subsequent evolution
is determined by the dynamics of brane collision. While in the case of
$D$-branes this dynamics is well understood (at least at the
level of an effective low-energy gauge theory), for a $D$-$\bar D$
system it is rather non-trivial.  For instance, it has been suggested
that the energy stored in the rolling tachyon can play the role of
dark matter\cite{darkmatter} However, it was shown recently in
\cite{kofmanlinde}, that unless the energy from the tachyon condensate
dissipates very efficiently, the tachyon will in fact overclose the
Universe.  Therefore, it is very important to understand the nature of
energy transfer from the tachyon to the ordinary particles.  Moreover,
stable $D$-branes of lower dimensionality, produced by annihilating
$D$-branes, can play the role of topological defects, such as cosmic
strings, monopoles, or domain walls, and can be potentially detected by
observations \cite{ST}. They may also help in liberating 
part of the tachyonic energy\cite{tachioncosm}.

The aim of the present paper is twofold. We point out some close
analogies between $D$ branes and field-theoretic topological defects.
We also identify some new decay mechanisms for defect-anti-defect
pairs and argue that similar mechanisms should operate in the decay of
$D -{\bar D}$ pairs.
The close analogy between $D$-branes and field theory topological
defects is interesting for two reasons.  First, it becomes possible to
model at least some aspects of $D$-brane dynamics on simpler field
theoretical models.  Second, since analogies are much closer than one
might naively expect, they may enable us to better understand the
solitonic nature of $D$-branes.

In the present paper, we shall explicitly demonstrate the following
properties.

 {\bf 1)} Annihilation of a pair of field theoretic topological defects (of
opposite topological charge) can produce stable lower dimensional 
topological defects. 

{\bf  2)} This annihilation can also be described in the language of
tachyon condensation. 

 {\bf 3)} Topological defects that support massless gauge field excitations
in their world-volume necessarily allow for the electric flux tubes
(open strings) ending on their world volume. They also imply the existence of
closed flux tubes (closed strings) in the bulk.  Thus, such solitons are analogous to
$D$-branes. 

{\bf 4)} The energy of annihilating defects is dissipated in the
form of strings and scalar waves.  In some exceptional models the
scalar wave production can be suppressed. Much as in the case of
unstable $D$-branes, one can construct classically-exact,
time-dependent localized lump solutions. These have the property that,
although the lump evolves in time, the energy stored in the original
soliton classically stays in the same plane and cannot dissipate into
scalar waves.

{\bf 5)} Nevertheless, such solutions are unstable with respect to the
resonant production of world-volume states, such as the localized
gauge fields.  Production of the world-volume gauge field (open string
states) during tachyon condensation is a generic feature of unstable
solitons that localize massless gauge field in their world volume.  We
expect similar effects to take place during the decay of unstable
$D$-brane systems.

{\bf 6)} Since the soliton localizes a massless gauge field, it
inevitably couples to the open and closed string sectors.  Both of
these states are produced during the tachyon condensation.  Closed
strings are expected to be the primary energy carriers in the bulk.
Such models share closest similarities with unstable $D$-branes as far
as energy dissipation is concerned.

{\bf 7)} When a pair of defects annihilates, the electric field
excited along the world-volume is squeezed into electric flux
tubes (strings), while the rest of the energy dissipates away.  The
strings form a random network consisting of infinite strings and
closed loops, and their subsequent evolution is
expected to be similar to that of
the conventional cosmic strings.

 \section{Defects Creating Lower Dimensional Defects.}

Topological defects are stable field configurations that appear in
theories with topologically non-trivial vacuum manifolds.  Most
commonly (but not necessarily) this happens when, as a result of the
Higgs effect, a certain symmetry group $G$ breaks spontaneously down
to its subgroup $H$.  Defects can be characterized by homotopy classes
$\pi_n(G/H)$, which classify uncontractible $n$-dimensional surfaces
in $G/H$.  Stable field configurations can exist whenever the
configuration of the order parameter (Higgs vacuum expectation value
(VEV)) $\Phi$ at space infinity is isomorphic to an uncontractible
$n$-surface in $G/H$. In a $D$-dimensional spacetime a stable defect
has $n+1$ transverse and $p=D - 2 - n $ longitudinal (world-volume)
space dimensions. We shall call such a defect a topological (or
solitonic) $p$-brane. In the simplest cases the VEV of the Higgs field
vanishes in the core of the defect, so that $G$ is restored.  However,
such a configuration may not necessarily be energetically favored, and
the actual structure of the core may be more complicated.  At the
moment we shall discuss the simplest case.

Topological $p$-branes are characterized by a topological charge $q_n$
corresponding to a given homotopy class of $n$-surfaces. In such a
case the two branes with opposite topological charge can annihilate
into the vacuum state. However, as noticed in \cite{tanmay},
$n$-homotopy topological charge may not be the only characteristic of
a $(D-2-n)$-brane. A given $p$-brane may be characterized by a pair of
topological charges $q_n, q_{n'}$ with respect to different homotopy
groups.  The existence of such a system can be understood as follows.
Let the vacuum manifold $G/H$ have two nontrivial homotopy classes
corresponding to $\pi_{n}$ and $\pi_{n'}$ respectively.  Let us assume
that $n < n'$. Then there are stable topological $(D-2-n)$-branes and
$(D - 2 - n')$-branes respectively, created by the same order
parameter $\Phi$. Each of these objects is stable due to a topological
obstruction at infinity. So if the two $p$-branes are separated far
enough in the transverse dimension, they continue to be stable.

Let us ask what happens if we try to place a $(D - 2 - n')$-brane with
the charge $q_n$ in the world volume of a $(D-2-n)$-brane with the
charge $q_{n'}$? When we do so, there is no topological reason for the $(D
- 2 - n')$-brane to be stable, and it will unwind.  This effect was
called ``defect erasure'' in \cite{tanmay}. A lower-dimensional defect
is erased by a higher-dimensional one. The existing topological charge
cannot disappear, however, and gets transfered to the higher dimensional
brane.  As a result, we shall get a $(D-2-n)$-brane with a pair of
homotopy charges $q_n,q_{n'}$. In other words, the $q_{n'}$ charge of
a $(D-2-n)$-brane is determined by how many $(D - 2 - n')$-branes it
has erased.

 Now, if such a  $(D-2-n)$-brane with charge $q_n,q_{n'}$ encounters
a  $(D-2-n)$-brane of charge  $-q_n, 0$, the result of the annihilation
will not be the vacuum state, but rather a stable  
 $(D - 2 - n')$-brane of charge $q_{n'}$. This is closely analogous to the
annihilation of string theory $D(p)$-branes. 

 We shall now study this effect in more detail on an explicit example
of \cite{tanmay}. This example considers the interaction between a magnetic
monopole and a domain wall. On the $D$-brane side it shares some
analogies with $D(p)$-$D(p+2)$ brane system studies by Gava, Narain and Sarmadi
\cite{Sarmadi}. 

We consider the $SU(5)$ Grand Unified model with an adjoint, $\Phi$,
(and a fundamental) scalar field. We take the standard Higgs potential
for $\Phi$,
$$
V(\Phi ) = - {1 \over 2} \, m^2  \, {\rm Tr}\Phi^2  \, + \,
                {h \over 4} \, ({\rm Tr}\Phi^2)^2  \, + \,
                {\lambda \over 4} \, {\rm Tr }\Phi^4.
$$
This potential has a  $Z_2$ symmetry: $\Phi \rightarrow
-\Phi$. 

The spontaneous
symmetry breaking 
\begin{equation}
SU(5)\times Z_2 \rightarrow 
[ SU(3)_c \times SU(2)_L \times U(1)_Y ]/Z_6 
\label{symbreak}
\end{equation}
occurs when $\Phi$ acquires a vacuum expectation value 
$$
\Phi_0 = {v \over {\sqrt{30}}} {\rm diag} (2,2,2,-3,-3) \ ,
\label{Y}
$$
where 
\begin{equation}
v = m \sqrt{{30} \over {30 h + 7 \lambda}} 
\equiv {m \over {\sqrt{\lambda '}}}\ .
\label{vev}
\end{equation}
To pick out this direction, we need the following constraints on 
the parameters in the Higgs potential 
$$
 \lambda > 0 \ , \ \ h > - {7 \over 30} \lambda \ \ 
(~ i.e.  \ \lambda ' > 0 ~ ) .
$$
The VEV $\Phi = -\Phi_0$ also leads to the symmetry breaking 
(\ref{symbreak}). The two discrete vacua $\Phi = \pm \Phi_0$ are 
degenerate due to the exact $Z_2$ symmetry.
It is well known that the symmetry breaking (\ref{symbreak})
leads to magnetic monopoles and $Z_2$ domain walls.

Following \cite{tanmay}, let us ask what happens when a monopole hits a
domain wall? Here, there
are two possibilities.
The first is when $\Phi =0$  and the other is when 
$\Phi \ne 0$ inside the wall. Both cases are possible, but here we
shall discuss only the first possibility. 

The most likely result of the monopole-wall encounter is that the
monopole will unwind on entering the wall where the full $SU(5)$
symmetry is restored.  The indications for that are as follows:

(i) Perturbatively, there is a short range
attractive force between the monopoles
and the walls coming from the Higgs exchange.
This is because the Higgs field inside the monopole,
as well as inside the domain wall, is not in the vacuum, which is costly in
Higgs energy. So a monopole can save the Higgs energy of its core
by entering the core of the wall where the zero Higgs VEV is 
supported by the wall. Thus, the monopole tends to form a bound state
with the wall \footnote{It was observed \cite{tanmay}
that the domain wall and
monopole bound state can lead to a classical realization of
a D-brane if the $SU(3)_c$ symmetry group further breaks to
$Z_3$ since now the monopoles bound to the walls will be
connected by strings. These strings will be the magnetic flux tubes
of  $SU(3)_c$ field. We will show below that if 
$SU(3)_c$ stays unbroken, there still are open strings ending on the
wall. These are the $SU(3)_c$-electric flux tubes.}. 

(ii) Once such a bound state is formed, as there is no topological
obstruction to the unwinding of monopoles on the wall, the monopoles
can continuously relax into the vacuum state.  Their magnetic charge
will then spread out along the wall.

(iii) Energetically, the most favored state is where the monopole
unwinds.

(iv)  Finally, in \cite{tanmay1} the interaction of domain walls and
monopoles was studied
numerically. It was found that monopoles indeed unwind when
passing through the wall.

The fundamental magnetic monopole in the above model
is essentially an SU(2) monopole
embedded in the full theory. 
The monopole solution has the following form:
\begin{equation}
\Phi_M \equiv \sum_{a=1}^{3} \Phi^a T^a + \Phi^4 T^4 +
\Phi^5 T^5 \ ,
\label{monopolesolution}
\end{equation}
where the subscript $M$ denotes the monopole field configuration,
$$
T^a = \frac{1}{2}{\rm diag}(\sigma^a,0,0,0)\ , \ \ 
T^4 = \frac{1}{2\sqrt{3}}(0,0,1,1,-2) \ ,
$$
$$
T^5 = \frac{1}{2\sqrt{15}}(-3,-3,2,2,2) \ ,
$$
$\sigma^a$ being the Pauli spin matrices, 
\begin{equation}
\Phi^a = P(r) x^a \ , \
\Phi^4 = M(r) \ , \
\Phi^5 = N(r) \ ,
\end{equation}
where $r=\sqrt{x^2+y^2+z^2}$ is the spherical radial coordinate.
The ansatz for the gauge fields of the monopole is: 
$$
W^a_i = \epsilon^a_{~ij} 
\frac{x^j}{er^2}(1-K(r))  \ , \ 
(a=1,2,3) \ ,
$$
\begin{equation}
W^b_i = 0, \ , \ \ (b \ne 1,2,3).
\label{gaugeansatz}
\end{equation}
The exact solution is known only in the case of a
vanishing potential (the BPS limit).
In the non-BPS case, the profile functions $P(r)$, $K(r)$, $M(r)$ and
$N(r)$ need to be found numerically. In \cite{tanmay2}  they  were
found by using a relaxation procedure with
the BPS solution serving as the initial guess.

The solution for a domain wall located in the xy-plane is
\begin{equation}
\Phi_{DW} = {\eta \over {2\sqrt{15}}} \tanh(\sigma z) (2,-3,2,2,-3) \ ,
\end{equation}
where $\sigma=\eta \sqrt{\lambda '/2}$.

When the monopole and the domain wall are very far from each other, 
the joint field configuration is given by the product ansatz:
\begin{equation}
\Phi = \tanh (\sigma (z-z_0)) \Phi_M \ ,
\label{initialfield}
\end{equation}
where $z_0$ is the position of the wall and $\Phi_M$ is the monopole
solution in eq. (\ref{monopolesolution}).

As we already mentioned, monopoles unwind when they are placed on top
of a domain wall, with their magnetic energy spreading along the wall.
To visualize this process, let us imagine the interaction of a
spherical domain wall and the monopole. For a large radius
($r >> \sigma^{-1}$) such a domain wall
can be approximated by
\begin{equation}
\Phi = \ {\eta \over {2\sqrt{15}}} 
{\rm tanh} (\sigma (\rho - r)) \ ,
\label{sphericalwall}
\end{equation}
where $\rho$ is a radial coordinate.
When such a domain wall erases a monopole, the magnetic charge of the monopole
gets uniformly distributed on the sphere. The field configuration
outside the sphere is then similar to the field 
of a magnetic monopole placed at
$\rho = 0$.  The spherical domain wall carries zero $q_o$ topological charge,
but nonzero $q_2$ magnetic charge.
Such a system is unstable and will collapse, producing a magnetic
monopole. 

If a monopole is erazed by an infinite planar domain wall, its
magnetic charge spreads along the wall in the form of Higgs and gauge
field excitations, but at any finite time it will be within a finite
radius of the point of the original encounter.  If the wall then
annihilates with an anti-wall, the magnetic charge will recollapse and
form a stable magnetic monopole.

Magnetic monopoles will be produced in wall annihilations even if the
colliding walls have no net magnetic charge.  This is due to quantum
(or thermal) fluctuations of the Higgs field and of the magnetic
charge density along the walls.  This mechanism of defect formation,
which is similar to the usual Kibble mechanism, will be analyzed in a
separate paper.  In the context of $D$-brane annihilation, it was
recently discussed by Sarangi and Tye \cite{ST}.

\section{Open Strings Ending on Domain walls}

\subsection{General Mechanism for Gauge Field Localization}

 It has been known for some time \cite{misha}
that domain walls, like $D$-branes \cite{Dgauge},
can support massless gauge fields in their world volume. It is interesting
that such a situation automatically implies that there must also be 
open strings ending on such domain walls. Strings in question are the
electric flux tubes. In addition, there must be
closed strings in the bulk. 

This result is very general and is independent of the detailed
mechanism of gauge field localization. Consider a domain wall (or any
solitonic brane) and assume that it localizes a massless Abelian gauge
field in its world-volume.  We shall assume that there is a mass gap
and no massless photon in the bulk theory.  This implies that test
electric charges located on the wall must be in the Abelian Coulomb
phase at low energies.  This can only happen if the test charges
outside the brane are in the confining phase. Indeed, if the bulk test
charges were in the Higgs phase, then the photon on the brane would be
massive, due to charge screening. If instead they were in the Coulomb
phase, then the gauge field could not be localized, simply by charge
conservation and universality.  Charges cannot go where there is no
massless photon. So the only regime that is compatible with the
charge-universality and flux conservation is the confining phase
outside the wall. This automatically implies that the electric field
outside the brane can only exist either in the form of flux tubes
attached to the wall (open strings), or the closed flux tubes (closed
strings) that can move freely in the bulk. This striking analogy with
the $D$-brane picture may indicate an intrinsic connection between the
existence of massless world-volume gauge field and of open strings in
the same theory.

In this section, we shall discuss this effect in more detail and
generalize it to unstable wall systems.  
We shall first review the general mechanism of Ref.~\cite{misha} and
later discuss the analogies with $D$-branes. We shall also show that
the $SU(5)$ domain walls discussed above localize massless gauge
fields by this mechanism. Due to this, there are automatically open
strings (QCD flux tubes) ending on such domain walls.

Following \cite{misha}, let us consider a toy model with
$SU(2)\otimes Z_2$ symmetry and two Higgs fields $\Phi$ and $\chi$.
$\chi$ is a real $SU(2)$-singlet field, which changes sign under the $Z_2$
symmetry. $\Phi$ is a $Z_2$-even $SU(2)$-triplet feld.
The Lagrangian of the system is
$$
{\cal L} = -\frac{1}{4g^2}G_{\mu\nu}^aG_{\mu\nu}^a + 
$$
\beq
\frac{1}{2}(D_\mu \Phi^a)^2 
- \frac{1}{2}\lambda' (\chi^2 + \kappa^2  - v^2 + \Phi^2 )^2 
+\frac{1}{2}(\partial_\mu\chi )^2
-\lambda(\chi^2 - v^2)^2\, ,
\label{Lagrstrc}
\eeq
where $G_{\mu\nu}^a $ is the gluon field strength tensor, $v$  and $\kappa$
are positive 
parameters having the dimension of mass and assumed  to be
much larger than the scale parameter $\Lambda$ of the $SU(2)$ gauge theory 
at hand
\footnote{Thus,  we  ignore
the shift of the vacuum energy due to the gluon condensate outside
the wall.}, $\lambda$ and  $\lambda '$ are (small) dimensionless coupling 
constants, and $g$ is the 
gauge coupling constant.

In the true vacuum of the theory, $Z_2$ symmetry is spontaneously 
broken, and the field $\chi$ develops a vacuum expectation value,
\beq
\chi = \, v \,\,\, \mbox{or} \,\,\, - \, v \, .
\label{2v}
\eeq
Correspondingly, the self-interaction potential for $\Phi$ is stable, and the
gauge $SU(2)$ is not spontaneously broken. The theory is in the confining 
phase. All observable degrees of freedom are bound states of
gluons and/or matter fermions (if such are included in the model), with masses 
of the order $\Lambda$. The mass of the $\chi$ quantum is  
\beq
m =\sqrt{2\lambda}\,  v \, .
\label{em}
\eeq
The theory has a stable domain wall interpolating between the two different 
vacua in Eq. (\ref{2v}),
\beq
\chi_0 = v \,{\rm tanh}\, (mz) \, .
\label{wall}
\eeq
 For definiteness  the wall is placed in the
$\{x,y\}$ plane; the width of the wall in the $z$ direction is
of order $m^{-1}$.

 Although $\Phi$ is zero in the vacuum, it can condense on the wall,
thereby breaking gauge symmetry.
To see that this is indeed the case, consider the behavior of $\Phi$ in the
classical wall background.  Consider 
a linearized equation for small perturbations in
$\Phi^a = \delta_{3a}\Phi_0{\rm e}^{-i\omega t}$
in the  kink background (\ref{wall}), 
\beq
\left\{ -\partial_z^2 + \lambda '\left[ \kappa^2 + v^2({\rm tanh}^2(mz) - 
1)\right]
\right \} \Phi_0 = \omega^2 \Phi_0\, . 
\label{arastab}
\eeq
A close examination of this equation shows \cite{misha} 
that in a wide range of the parameter space, 
$\Phi$ becomes tachyonic and condenses in the core of the defect.
Thus, inside the wall
the $SU(2)$ gauge symmetry is spontaneously broken down to $U(1)$.
Two out of three gluons acquire very large masses of order of $v$ 
in the vicinity of the wall.
The third gluon becomes a photon. 

The phase portrait of the theory emerging in this way is the
following.  Outside the wall the theory has a wider gauge invariance,
$SU(2)$, and is in the non-Abelian confining phase.  The $U(1)$ gauge
invariance is maintained everywhere -- inside and outside the
wall. The test charges inside the wall interact through the photon
exchange.  The theory inside the wall is in the Abelian Coulomb phase.
The photon and the test charges cannot escape in the outside space
because there they become a part of the confining $SU(2)$ theory with
no states lighter than $\Lambda$.  The three-dimensional low-energy
observer confined inside the wall sees a massless photon.

\subsection{Flux Spreading in the Wall}

Let us now study in more detail how the Coulomb phase appeares in the
world volume theory of the wall.  As discussed above, the key point is
that the non-Abelian theory is in a partially-Higgs phase inside the
wall, and confinement cannot penetrate there.  That is, the wall plays
the role of a vacuum layer inside a dual superconductor.  The
non-abelian fields are repelled away from the wall by a dual Meissner
effect.  In order to study this dynamics more explicitly, it is useful
to have a model in which the thickness of the non-confining layer
could be arbitrarily changed. Let us introduce a model of this
sort. We shall see that at the same time the example gives an explicit
realization of an unstable brane, which supports a massless gauge
field in the world-volume.

Our task is rather simple. The general mechanism of \cite{misha} tells
us that whenever we create a layer of a non-confining phase in the
confining vacuum, a massless $U(1)$ gauge field will be trapped in
the layer.  Thus all we need is to find a theory that could supports
both confining and Coulomb vacua.  In order to do this, let us
modify the model (\ref{Lagrstrc}) in such a way that one of the
degenerate vacua is at $\Phi = 0$, while the other is at $\Phi \neq 0$.
This will enable us to construct a stable wall
interpolating between the confining and Coulomb phases. Next, by taking a
wall-anti-wall pair we can realize an unstable lump supporting a
$(2+1)$-dimensional Coulomb phase embedded in a $(3 + 1)$-dimensional
confining space.

In order to achieve this goal, we have to modify the Higgs potential of
the $SU(2)$ model of \cite{misha}. The needed potential
is provided by an $N=1$ supersymmetric theory with the superpotential
\begin{equation}
 W = {{\rm Tr}\Phi^3\over 3} + {\rm Tr}\Phi^2 \chi - \chi\mu^2 +
{\chi^3 \over 3}.
\label{superpotential}
\end{equation}
The scalar fields of the previous model are promoted to chiral
superfields. Note that the $Z_2$ symmetry of the previous example now
becomes an $R$-symmetry, under which 
\begin{equation}
 W \rightarrow - W, ~~ \Phi  \rightarrow - \Phi,~~~
 \chi \rightarrow -\chi.
\label{Rsymmetry}
\end{equation}
Transformation properties under the gauge $SU(2)$ are unchanged.
The vacua of the theory can be found by solving the standard $F$-flatness
and $D$-flatness conditions. The latter one,
\begin{equation}                        
\left [\Phi, \Phi^*\right ] = 0,
\label{dflat}
\end{equation}
suggests that $\Phi$ can be brought to a diagonal form
$\Phi = {\phi \over \sqrt{2}} {\rm diag}(1, -1)$ by a gauge transformation.
With this form, the only remaining vacuum conditions are
\begin{equation}
W_{\chi} =  \phi^2 - \mu^2 + \chi^2 = 0
\label{f1}
\end{equation}
 and
\begin{equation}
W_{\phi} =  2\phi\chi = 0.
\label{f}
\end{equation}

There are four degenerate vacuum states $(\phi = \pm \mu, \chi = 0)$
and $(\chi = \pm \mu, \phi = 0)$. There will be corresponding domain walls.
We are interested in domain walls that interpolate between
$\phi =\mu $ and $\phi = 0$ vacua. Notice that the value 
of the superpotential is different on different sides of such walls,
\begin{equation}
W(+ \infty) - W(- \infty) =  {2\over 3} \mu^3.
\label{centralcharge}
\end{equation}
On such a background, $N=1$ supersymmetric algebra
admits a central extension with the central charge 
given by (\ref{centralcharge})\cite{misha,misha1}.
This implies the existence of 
BPS saturated walls (see Appendix). Indeed, the BPS conditions
\begin{equation}
\partial_z\chi \pm  (\phi^2 - \mu^2 + \chi^2) = 0
\label{BPS1}
\end{equation}
 and
\begin{equation}
\partial_z\phi \pm 2\phi\chi = 0
\label{BPS2}
\end{equation}
can be easily solved, yielding exact solutions for the domain wall
profile ($z$ is the coordinate perpendicular to the wall).  
Choosing the upper signs in Eqs.(\ref{BPS1}),(\ref{BPS2}), we have
\begin{equation}
\phi =  {\mu \over 2} \left ({\rm tanh}(z\mu) + 1\right )  
\label{BPSW1}
\end{equation}
 and
\begin{equation}
\chi =  {\mu \over 2} \left ({\rm tanh}(z\mu) - 1\right )  
\label{BPSW2}
\end{equation}

The domain wall divides space into two regions. $z < 0$ is the region
in which the full $SU(2)$ is unbroken and perturbatively all three
gauge bosons $(W^{\pm}_{\mu},A_{\mu})$ are massless.  But below
certain scale $\Lambda$, the $SU(2)$-gauge theory is strongly coupled and
is in the confinement phase. The lowest-energy degrees of freedom are glueballs of $SU(2)$ of mass
$\sim \Lambda$, and there are no massless excitations in the
spectrum. We shall assume that $\Lambda << \mu$.  In the $z > 0$
domain, $SU(2)$ is broken to $U(1)$ at the scale $\mu$, and the theory is
in a partially Higgs phase. Two $(W^{\pm}_{\mu})$ out of three gauge
fields gain masses $\sim \mu$. As a result, the only degree of freedom
below this scale is a massless photon $A_{\mu}$, and the theory is in the
Abelian Coulomb phase. Ignoring confinement, we can find the low energy
perturbative spectrum by solving the linearized equation for the gauge
fields about such a background. For the charged gauge fields this
equation has the following form:
\begin{equation}
\left (\partial^2 -  g^2{\mu \over 4}^2 \left ({\rm tanh}(z\mu) + 1\right )^2
\right )W_{\mu}^{\pm} = 0  
\label{perturbative}
\end{equation}

This is just a Schrodinger equation with an infinitely wide barrier.
Modes with $z$-momentum lower than $\mu$ cannot propagate in the $z > 0$ 
domain and have exponentially suppressed wavefunctions,
\begin{equation}
 \Psi (z) \sim {\rm exp(-g\mu z)}.
\end{equation}
This is nothing but the non-Abelian Meissner effect. On the other hand,
for the photon the barrier is transparent. Thus, the test charges localized
in $z> 0$ domain can effectively interact only via exchange of photons.
Now let us take into account the effect of the confinement in $z < 0$
domain. Since there is a mass gap, the low energy photon cannot
penetrate in this domain. However, since the effective low energy
theory in $z > 0$ domain is an unbroken $U(1)$, an observer in this
domain must see a massless photon. 
Thus, there is a massless photon, but its wave function gets exponentially
suppressed in the left domain, due to dual Meissner effect.
Charges in the right domain continue to be in the Coulomb phase,
but the photon flux gets repelled from the left domain. 
 
If we try to move a test charge from the right domain into
the left one, the tube of the photon electric 
flux will be stretched between the
charge and the wall. The throat of the tube will open up in the right
domain where flux can freely spread out.  The troat of the tube will
be seen by inner observers 
as the charge of the same magnitude that was taken out.  Thus, an observer in this domain will
see charge conservation.

 Now it is not hard to see what happens if an anti-wall is
placed parallel to the wall at some $ z = \Delta >0$. This anti-wall
takes $\phi$ back to the trivial vacuum $\phi = 0$ with the confinement
and the mass gap. For
$\Delta >> \mu^{-1}$, the wall-anti-wall configuration can be approximated
by the product ansatz:
\begin{equation}
\phi =  {\mu \over 4} \left ({\rm tanh}((z - \Delta)\mu) + 1\right ) 
\left (-{\rm tanh}(z\mu) + 1\right ).
\label{wallantiwall!}
\end{equation}
Now the photon wave function is repelled from both sides, and the
photon is trapped in the layer where $SU(2)$ is in the partially-Higgs phase.
This layer is nothing but an unstable non-BPS brane, which supports
a massless photon in its world volume.  We can integrate out all the heavy degrees of freedom
and write down an effective low energy theory. This theory consists of only  $U(1)$ gauge field,
with the gauge-kinetic function of the following form
\begin{equation}
 \psi(z) \, F_{\alpha\beta} \, F^{\alpha\beta}  \, + \, {\tilde\psi}(z) \, F_{\alpha z}
 \, F^{\alpha z} \, + \, {\rm high-derivative ~  operators ~ scaled ~  by} \Lambda,
\label{kinetic}
\end{equation}
where $F_{\alpha\beta} \, = \, \partial_{\alpha} \,  A_{\beta} \,  - \,  
\partial_{\beta} \,  A_{\alpha}$, the indices 
 $\alpha, \, \beta \, = \,  0, \, 1, \, 2$ correspond to the world-volume 
coordinates, and $\psi(z)$ and ${\tilde\psi}(z)$ are localized function 
of width $\Delta$.  Note that they
are in general different due to the spontaneous breaking of four-dimensional 
translational invariance by the wall.  Since the localization is due to the
bulk confinement effects, the exact form of the functions $\psi(z), \, 
{\tilde\psi}(z)$ is unknown, but they must fall off exponentially fast,
\begin{equation}
\psi(|z| \gg \delta) \, \sim \,  {\tilde\psi}(|z| \gg \delta) \, \sim \, 
{\rm e}^{-k\Lambda |z|}
\end{equation}
with $k\sim 1$.
This fall-off is simply a consequence of the fact that the light gauge 
degrees of freedom cannot penetrate in the bulk, where there are no states 
lighter than $\Lambda$.
Hence, all the correlation functions must fall off exponentially fast. 
The only behaviour compatible with this fact and with the unbroken 
$U(1)$-gauge invariance, is that shape functions $\psi$, $\tilde\psi$ 
must die-off, as given above.
Then it is obvious that $A_{\alpha}$ has a  $z$-independent zero-mode 
component that satisfies
the three-dimensional massless equation
\begin{equation}
\nabla^2_{2+1} a_{\alpha}(x) = 0.
\label{masslessa }
\end{equation}
This component is normalizable due to the finiteness of the following integral
\begin{equation}
\int \, dz \,  \psi(z).
\end{equation}
Consequently, there  are open strings ending
on the wall; these are the electric flux tubes of the confining bulk theory.
When wall-anti-wall annihilate, the remaining vacuum will have
no open string excitations, but only closed strings (glueballs of
the bulk $SU(2)$).

Again, this effect is in full analogy with the $D(p)$-brane picture.  If
this analogy is indeed deep, it may sheds some light on the $D$-brane
dynamics. For instance, in the $D$-brane picture, all the open string
modes must decouple in the tachyonic vacuum.  Here, we see this
effect explicitly.  In this language, it is obvious why there are no
tachyonic modes in the gauge field, but only a zero mode. When branes
annihilate, the wave function of the photon simply vanishes, which
signals that the description in terms of a massless photon is no
longer possible. Photon becomes non-dynamical and the only relevant
degrees of freedom are glueballs (closed strings) of the strongly 
coupled confining theory.

 Moreover, once the brane becomes thinner than the penetration length,
$\Delta << \mu^{-1}$, it can no longer support the spread-out of the 
flux.  At this point, there are only closed (or infinitely long) strings 
in the theory.

\subsection{Spread-out of Flux in the Dual Example}.

The spread-out of flux in the world-volume of the layer is the
key ingredient for establishing the Coulomb phase there. Dynamics of this
spread-out can be best understood on a toy "dual" example, in which 
the roles of magnetic and electric charges are interchanged. 
That is, now the magnetic charges are in the Coulomb phase on the
brane and in the confining phase in the bulk. Let us construct such a dual
example. In fact all we have to do is to take the previous example
 and substitute $SU(2)$ by $U(1)$, and adjoint $\Phi$ by a pair
of singlet fields $\Phi, \bar{\Phi}$, with opposite $U(1)$-charges.
The superpotential now becomes:
\begin{equation}
 W =  \bar{\Phi}\Phi \chi - \chi\mu^2 + {\chi^3 \over 3}.
\label{superpotentialdual}
\end{equation}
Much in the same way, this model also has a 
$Z_2$ $R$-symmetry 
\begin{equation}
 W \rightarrow - W, ~~ \Phi,\bar{\Phi}  \rightarrow - \Phi,\bar{\Phi},~~~
 \chi \rightarrow -\chi.
\label{Rsymmetrydual}
\end{equation}
The four degenerate vacuum states are $(\Phi = \bar{\Phi}=\pm \mu,
\chi = 0)$
and $(\chi = \pm \mu, \Phi = 0)$. 
The domain walls of our interest are the ones that interpolate between
$\Phi =\mu $ and $\Phi = 0$ vacua. These walls are again BPS states,
due to the existence of the central charge,
\begin{equation}
W(+ \infty) - W(- \infty) =  {2\over 3} \mu^3.
\label{centralchargedual}
\end{equation}
The wall solutions can be explicitly found and are similar to the
ones in the previous model,
\begin{equation}
\Phi = \bar{\Phi}= {\mu \over 2} \left (-{\rm tanh}(z\mu) + 1\right )  
\label{BPSdual1}
\end{equation}
 and
\begin{equation}
\chi =  {\mu \over 2} \left (-{\rm tanh}(z\mu) - 1\right ).  
\label{BPSdual2}
\end{equation}
Now, in the $z < 0$ domain $U(1)$ is in the Higgs phase and there are
topologically stable magnetic flux tubes (cosmic strings). These strings can
end on the wall, since the flux can freely spread out in the right domain,
where $U(1)$ is unbroken. Magnetic charges are in the Coulomb phase
everywhere in this domain, but electric charges are partially screened close to
the boundary, due to presence of the infinite superconductor.
Let us again consider an anti-wall
placed parallel to the wall at some $ z = \Delta$ plane. This anti-wall
takes $\bar{\Phi},\Phi$,  back to the Higgs phase.
For $\Delta >> \mu^{-1}$ the wall-anti-wall configuration can be approximated
by the product ansatz:
\begin{equation}
\bar{\Phi}= \Phi = {\mu \over 4} \left (-{\rm tanh}
((z - \Delta)\mu) + 1\right ) 
\left ({\rm tanh}(z\mu) + 1\right )
\label{wallantiwalldual!!}
\end{equation}

Thus, we have created a layer of vacuum in an infinite superconductor.
This layer is our non-BPS brane, and is unstable. However, we want to
know whether magnetic charges in the layer are in the Coulomb phase.
To answer this question, consider the following experiment. Take an infinite
straight magnetic flux tube parallel to the layer. What happens if the tube
enters the layer? Will the magnetic flux spread uniformly or stay in the
tube?
 This question is simplest to answer for the cylindrical layer. That is,
consider
a cylindrical layer created by a cylindric wall- anti-wall
system concentrically embedded
inside one another. Inside and outside of the cylindrical layer, the
theory is in the Higgs phase. Let us assume that the 
cylinder is extended in the $z$
direction. If the radius of the inner wall $R$, as well as the wall-anti-wall
separation $\Delta$, is large, the configuration can be approximated by
the following function
\begin{equation}
\bar{\Phi}= \Phi = {\mu \over 4} \left (-{\rm tanh}(\mu(r - R)) + 1\right ) 
\left ({\rm tanh}(\mu(r-(R + \Delta)) + 1\right ).
\label{wallantiwallcylinder!!!}
\end{equation}
This configuration is unstable for two reasons. First, wall and anti-wall
tend to attract and annihilate. Second, each of them individually wants
to collapse under the tension force. However, if $R >>\Delta>> \mu^{-1}$,
we can ignore the time evolution. Now, let us see what will happen with
the straight infinite string after entering the layer. First off all,
notice that there is a short range attractive force between the 
string and the layer, due to the Higgs energy:
inside the layer the Higgs VEV vanishes, and thus the string will cost no
additional zeros of the Higgs field, whereas outside the very 
existence of strings
requires extra zeros, which are costly in energy.
This is due to the same reason as the attractive
force between monopole and domain wall in the $SU(5)$ example. Thus, the
string will form a bound state with the layer.  Once inside the layer,
there is no topological obstruction for the flux to spread uniformly
around the cylinder. Whether this will happen is now a dynamical
question. To answer it, it is useful to first consider the case of a
global string.

{\bf Global Flux Spread-out}

Assume for a moment that the $U(1)$ symmetry is not gauged. Then strings in
question are global strings. In such a case, there is no gauge field to
compensate gradient energy at infinity, and the string energy
per unit length diverges logarithmically in the $r$ direction.
If the flux stays localized in the tube, then it's energy will effectively
be equal to the one of the global string with an effective short
distance cutoff at $r\sim \Delta$.  That is \cite{vilenkinglobal},
\begin{equation}
E_{localized} \sim \mu^2 {\rm ln}(r_{max}/\Delta),
\label{energygloballocalized}
\end{equation}
where $r_{max}$ is the large distance cutoff. On the other hand,
for the uniformly spread flux the energy is
\begin{equation}
E_{localized} \sim \mu^2 {\rm ln}(r_{max}/R)
\label{energyglobalspead}
\end{equation}
and is by $\sim \mu^2 {\rm ln}(R/\Delta)$ smaller. So the flux will spread.

{\bf Magnetic Flux Spread-out}

 Let us now come back to the gauged $U(1)$. Now the gradient energy
of the phase is compensated by the gauge field, which assumes a pure gauge
form
\begin{equation}
 A_{\theta} = n,
\label{gaugefield}
\end{equation}
where $n$ is the winding number. This compensation occurs at the expense of
the magnetic energy. The reason is that the gauge field must vanish
at the origin and cannot be pure gauge everywhere. This results in
a finite magnetic energy $E_{magnetic} \sim \mu^2$.
Now let us estimate the magnetic energy of the configuration when the
flux is uniformly spread around the cylinder. Since $\Phi$ vanishes
at the layer, $A_{\theta}$ can be zero everywhere in the inner domain
and be  $A_{\theta} = n$ outside. Thus  $A_{\theta}(r)$ changes
from $0$ to $n$ in a small interval $\sim \Delta$.\footnote{
Strictly speaking in the configuration described by 
(\ref{wallantiwallcylinder!!!}) $\Phi$ is nowhere zero, but
is exponentially small. For $A_{\theta}$ to vanish in the inner domain,
$\Phi$ must become strictly zero on some circle. This will only cost
exponentially small energy, as we shall see,
negligible with respect to the relatively enormous gain
due to resulting spread-out of the magnetic flux.}
The resulting magnetic energy of such a configuration is
\begin{equation}
E_{magnetic} \sim 2\pi \int_0^{\infty} {dr \over r}
(\partial_r  A_{\theta})^2  \sim n^2\mu^2 {\rm ln}(1 + \Delta/R),
\end{equation}
which vanishes for $R \rightarrow \infty$. This indicates that the magnetic
flux would prefer to spread. Thus, magnetic charges inside the layer
will be in the $(2+1)$-dimensional Coulomb phase.

\section{Brane Annihilation and Energy Dissipation in the Bulk}

 When $D(p)-\bar{D}(p)$ system annihilates, its energy can dissipate in the
bulk in the form of closed strings (or lower dimensional D-branes).
The same is true for solitons that support massless gauge fields
in their world volume. Indeed, according to the mechanism of
\cite{misha}, in order for a massless gauge field
to appear in the world volume of a domain 
wall, the electric charges must be in the
confining phase in the bulk. The bulk theory then has a mass gap
\footnote{We assume there are no massless ``pions'' in the bulk.},
and the lowest mass excitations are closed strings.
These are the ``glueballs'' of the confining bulk theory.
There is an intrinsic connection between the existence of a massless
photon on the brane and closed strings in the bulk.
These strings are the lightest agents that
can carry away the energy of the unstable wall configurations.

It has been suggested that in $D$-brane annihilation, at least in the
limit of decoupled closed strings, the energy classically stays
localized, although the tachyon evolves in time.  
This behavior is very different from that of annihilating solitons.
The difference could be attributed to the fact that
solitons are
made of a scalar field, which is a bulk mode. Thus, there are
excitations of the scalar field (e.g., plane waves) about the bulk
vacuum which can carry away the energy of the annihilating solitons,
while there are no obvious analogous constituents for $D$-branes.

Below we shall discuss solitonic
solutions with similar properties.  Much as in the
$D$-brane case, in these solutions the tachyon evolves in time, but 
classically the energy density stays localized.  
Although the solitons in question
are made of a scalar field, the quanta of this field are
infinitely heavy, and disipation into scalar waves is suppressed.
Nevertheless, we show that there is a universal source of instability,
which is due to the resonant production of the localized world-volume
gauge field. This mechanism is largely insensitive to the detailed
structure of the soliton, and 
we expect that a similar effect should take place for $D$-branes.

We shall now discuss this issue in more detail.   
In some models, a wall-anti-wall pair decays
predominantly into Higgs particles, even though they are much more
massive than glueballs.  Annihilation of kinks and anti-kinks in the
$(1+1)$-dimensional $\phi^4$ theory has been extensively studied
(see, e.g., \cite{kinkan} and references therein) 
and reveals surprisingly rich physics.  Depending on the
impact velocity, the kink and anti-kink are either reflected from one
another, or they form a relatively long-lived bound state.  In the
latter case, the bound pair oscillates and gradually dissolves by
radiating $\phi$-waves.  In the models of interest to us, $\phi$ is
coupled to gauge fields, and this oscillating $\phi$ background will
produce pairs of gauge quanta.  The energy of the quanta will be
determined by the frequency of the oscillation, which is typically
comparable to the scalar field mass $m$.  If $m$ is large, the initial
energy of the gauge quanta can be much greater than the confinement
scale $\Lambda$, but the quanta will gradually degenerate into
qlueballs.  Thus, the energy of the annihilating walls is carried away
by $\phi$-particles and by closed string excitations.  The fraction of
energy dissipated into each channel is model-dependent.

Here, we shall
be interested in models where the classical soliton decay into 
scalar waves is suppressed and can be ignored in the first
approximation.  To this end,
we can consider the extreme case, in which the Higgs field
excitations
in the bulk are infinitely massive, although the domain walls
and strings have a finite tension. 
 
An explicit model of this type can be easily constructed. We take
$N=1$ supersymmetric $SU(2)$ gauge theory with one chiral superfield
$\Phi$ in the adjoint representation. We choose the superpotential in the form
\begin{equation}
 W = {1 \over 4} {\rm Tr}\Phi^2 ({\rm ln}({\rm Tr}\Phi^2) - 1)
\label{wlog}
\end{equation}
The above equation is written in units of some fundamental mass scale.
This theory has two degenerate vacuum states: 
$(1)$ an $SU(2)$-invariant vacuum $\Phi = 0$,  and $(2)$ a vacuum with
$\Phi = {1 \over \sqrt{2}} (1, -1)$, where $SU(2)$ is broken to $U(1)$.
(As usual, all non-diagonal
components can be set to zero due to $D$-flatness conditions).
Let us investigate both of these vacuum states. In the $SU(2)$-invariant
vacuum, the mass of $\Phi$-quanta is infinite. So the low energy theory
is a pure supersymmetric $SU(2)$-gluodynamics. It is well known
\cite{witten} that
this theory exhibits confinement and a mass gap at a scale $\Lambda << 1$.
There is an anomalous $U(1)_R$ chiral global symmetry, which is 
broken down to $Z_2$ by instantons.
Perturbatively, the value of the superpotential is zero in this vacuum.
Non-perturbatively, there is a low-energy superpotential generated 
though the gaugino condensation \cite{gaugino},
\begin{equation}
 W_{dynamical} = \langle \bar{\lambda}\lambda \rangle \sim \Lambda^3.
\label{gauginow}
\end{equation}
This condensate spontaneously breaks the anomaly-free $Z_2$. The low energy
degrees of freedom in this vacuum are glueballs (or closed strings)
of mass $\sim \Lambda$.

 In the  $\Phi = {1 \over \sqrt{2}} (1, -1)$ vacuum, $\Phi$-quanta and
the two out of three $SU(2)$ gauge superfields have masses $\sim 1$. The only
massless degree of freedom is the photon and the theory is in 
the Abelian Coulomb phase.

 Due to the vacuum degeneracy, there are domain walls that 
interpolate between $SU(2)$-confining and $U(1)$-Coulomb vacua\footnote{
Note that in addition there are BPS domain walls interpolating between the
two degenerate confining vacua, due to
spontaneous breaking of $Z_2$ \cite{misha,  qcdwall}.
However, we shall only be interested in the walls that interpolate between 
confining and Coulomb phases.}.
 These walls satisfy the BPS equation
\begin{equation}
\partial_z \phi + {\phi\over 2}{\rm ln}\phi^2 = 0,
\end{equation}
where 
\begin{equation}
\Phi = {\phi \over \sqrt{2}} (1, -1).
\label{defphi}
\end{equation}
This can be explicitly integrated and gives
\begin{equation}
\phi_w(z) = {\rm exp}(-e^{-z}).
\label{BPSlog}
\end{equation}
The tension of the wall is given by the difference of the superpotential VEVs
on the two sides of the wall and is perturbatively equal to
\begin{equation}
T = W(-\infty) - W(+\infty) = {1 \over 4}.
\label{1/4}
\end{equation}
Non-perturbative effects due to gaugino condensation in the confining vacuum
introduce negligible corrections $\sim \Lambda^3 << 1/4$. Since the wall is the
boundary between confining and Coulomb phases, there are flux tubes that
end on it. We are interested in an unstable layer of Coulomb vacuum 
bounded by two confining regions. As discussed above, there is a
massless photon localized within the world volume theory of the layer.
Such a layer is formed by a wall-anti-wall
system and is unstable, since wall and anti-wall tend to annihilate.
The energy stored in the walls will be released in the form of 
closed string excitations. 
 
The model (\ref{wlog}) was constructed so that in the $\Phi = 0$
vacuum the Higgs field has an infinite mass, and thus has no harmonic
plane wave excitations in the bulk.  Nevertheless, there are other
types of waves.  For $\Phi$ of the form (\ref{defphi}), the field equation is 
\begin{equation}
\partial^2\phi + V'(\phi) = 0,
\end{equation}
where
\begin{equation}
V(\phi) = |W'(\phi)|^2 = {1\over{4}}\phi^2\ln^2(\phi^2).
\end{equation}
For homogeneous, small-amplitude oscillations, this has an approximate
solution
\begin{equation}
\phi(t)\approx A\sin [m(A)t]
\label{approxwave}
\end{equation}
with
\begin{equation}
m(A)=\sqrt{2} |\ln A|.
\label{mA}
\end{equation}
This is accurate as long as $(\ln A)^2\gg 1$.  

Travelling wave solutions can be obtained from (\ref{approxwave}) by
applying a Lorentz boost.  They have the form of the usual plane waves
for a massive particle, the only unusual feature being that the
particle's mass depends on the amplitude $A$ of the wave.  The effect of
this feature on wave propagation is not difficult to figure out.  For
annihilating defects of codimension $(n+1)$, the amplitude of massless
radiation decreases with the distance as $r^{-n/2}$, and for usual
massive particles there is an additional decrease due to the spreading
of the wave packets (this is the only effect in the case of
codimension-1 defects).  In our model, as the amplitude of the waves
decreases, the effective Higgs mass (\ref{mA}) will grow, and the
speed of the waves will decrease.  Asymptotically, the propagation
speed approaches zero, but in our model this slowdown process is very
slow, since the mass dependence on $A$ is only logarithmic.  As in the
$\phi^4$ model, closed string states will also be produced, but again,
the relative efficiency of the two channels is hard to assess.

If one wants to suppress the scalar radiation altogether, one has to
construct a model admitting non-dissipative ``breather'' solutions.
One example is the sine-Gordon model, where the breather describes an
oscillating kink-anti-kink pair.  Another interesting example is the
(non-supersymmetric) model with the potential
\begin{equation}
V(\Phi) = -{{\rm Tr}\Phi^2\over 4} {\rm ln}( {\rm Tr}\Phi^2).
\label{layer}
\end{equation}
Such a potential was first considered by Minahan and Zwiebach in 
\cite{lump} as a model
for an unstable brane solution. (Note, however, that 
our treatment of the gauge fields
is very different.) This model has an unstable (but static) lump solution 
\cite{lump}
\begin{equation}
\phi_{lump}(z) = e^{-{z^2 \over 4}}.
\end{equation}
$\phi$ vanishes outside the lump, and the mass of $\Phi$ is infinite.
Since in our case $\Phi$ is an adjoint field, the theory is in the
confining phase outside, and in the Coulomb phase inside the lump, and the lump
supports a massless photon in its world-volume. 
The lump is unstable, however, and will decay.

Remarkably, there is a (perturbatively) exact time-dependent
solution, with a localized energy density:
\begin{equation}
\phi(z,t) = A(t)\phi_{lump}(z),
\label{oscillator}
\end{equation}
where $A(t)$ satisfies the oscillatory equation
\begin{equation}
{d^2A(t) \over dt^2} = {1\over{2}}A(t){\rm ln}( A^2(t)).
\label{amplitude}
\end{equation}
Although the lump vibrates locally, the energy
density does not dissipate and stays localized in the plane of the lump.

To investigate the stability of this vibrating lump solution, 
we consider linearized
scalar perturbation on this background. These perturbations satisfy the
following equation:
\begin{equation}
\left [ \nabla^2 + \left ( {z^2 \over 4} - {3\over 2} - 
{{\rm ln}A(t)^2 \over 2} \right )\right] \delta\phi = 0
\label{vibrations}
\end{equation}
For homogeneous perturbations, after the separation of variables
\begin{equation}
\delta\phi(t,z) = f(t)\rho(z),
\end{equation}
we find that $\rho(z)$ and $f(t)$ satisfy
\begin{equation}
\left [ - \, d_z^2 \, + \, \left ( \, {z^2 \over 4} \,
 - \, {3\over 2} \,  
 \right )\right] \,\rho \, =  \, m^2 \,\rho
\label{rhoeq}
\end{equation}
and 
\begin{equation}
\left [ \, d_t^2 \,  -  \, {{\rm ln}A(t)^2 \over 2}\right] \, 
f(t) \, = \, - \, m^2 \, f(t).
\label{feq}
\end{equation}
Equation (\ref{rhoeq}) has a single negative eigenvalue solution
$m^2=-1$, which
corresponds to the time-dependence of the vibrating background itself. There
is also a zero mode $m^2 = 0$,
\begin{equation}
\delta\phi = \, - \,  e^{-{z^2 \over 4}}\, {z \over 2} \, f_0(t),
\end{equation}
where $f_0(t)$ is given either by
\begin{equation}
f_0(t) \, = \, A(t)
\label{ fA}
\end{equation}
or by
\begin{equation}
f_0(t) \, = \, A(t) \, \int \, {dt' \over A^2(t')}
\label{ fA1}
\end{equation}
The rest of the eigenvalues are positive. This indicates that there
are no other tachyonic small perturbations in $\phi$.  That is, all
the modes are oscillatory about the background.  Thus, the classical
decay of this system into scalar waves is suppressed.  

These
arguments, however, do not take into account the possible parametric
resonance effects.  Although there are no homogeneous 
imaginary-frequency modes,
the amplitude of some of the positive-frequency modes can
grow due to paramertic resonance amplification. To see that such an
effects may indeed take place, we consider 
perturbations of finite wavelength and write
\begin{equation}
\delta\phi(t,z,\vec{x}) = f(t,\vec{x})\rho(z)
\end{equation}
where $\vec{x}$ are world-volume space coordinates. 
Taking the Fourier transform with respect to $\vec{x}$, we obtain the 
following equation for $f(t,\vec{k})$
\begin{equation}
\left [ d_t^2 \,   +  \,  (k^2 \,  +  \,  m^2  \, -  \, {{\rm ln}A(t)^2 
\over 2} \, ) \, \right] \, f(t,\vec{k}) \,  = \,  0,
\label{feq1}
\end{equation}
while $\rho(z)$ still satisfies Eq. (\ref{rhoeq}).
This is an equation of an oscillator with a time-dependent mass
term. The time dependence is periodic through the function $A(t)$. It
is well known that such systems exhibit parametric resonance effects.
That is, for some values of the parameter the amplitude grows
exponentially.  For a small amplitude, this happens when the average
value of the mass is approximately equal to half integer times the
frequency of mass-oscillation.  For large amplitudes, the
resonance bands become wider.  Based on this analogy, the above
equation is expected to exhibit parametric resonance for certain
resonant values of $k^2 \, + \, m^2$. The situation is somewhat
complicated by the fact that ${{\rm ln}A(t)^2 \over 2}$ blows up every
time $A$ passes through zero. However, $A$ spends near zero a very
small portion of the full period, and for an approximate analysis one can
smooth out the singularity.

Another source of instability is due to the
bulk gauge fields.  Coupling to these fields is crucial for
localizing the gauge field on the lump, and thus cannot be neglected.
In other words, there is no limit in which we can ignore coupling to
the bulk confining theory without undoing the localization of the
world-volume photon.  Therefore, even though production of scalar
waves is suppressed in this particular example, the energy density of
the vibrating lump will still dissipate in the form of glueballs
(closed strings) of the confining theory.  This dissipation can go in
two channels.  One is the direct production of $W^{\pm}$ bosons that
get localized masses from the lump field.  When the lump vibrates, so
does the effective mass of these particles, and this leads to particle
creation.  However, there is also a second channel, which is more
efficient. As we show below, the vibrating lump causes
a resonant amplification of the localized $U(1)$ gauge field, that is,
of open string states.  These states later annihilate into the bulk
glueballs, or survive in the form of long flux tubes, similar to cosmic 
strings. 

Before proceeding, we remark that non-dissipative 
breather solutions are not known
for defects with codimension greater than one, and we expect that
radiation of scalar waves cannot be avoided when such defects
annihilate.

\section{Production of World-Volume Gauge Fields (Open Strings)}

We now want to show that the tachyon condensation in the above model
leads to the production of localized gauge field quanta, that is, of 
the open string
states. This effect is generic for any unstable solitons that localize
a massless gauge field via the mechanism of bulk confinement, and we
expect it should also persist for $D$-branes.  Thus, we expect that
open string states are produced during the tachyon condensation in an
unstable $D$-brane decay.  We shall first review the generic mechanism
of the production and later discuss an explicit model.  

The effective low-energy action for the zero mode world-volume photon
$a_{\alpha}(x^{\beta})$ is the following:
\begin{equation}
\int  d^3xdz \, \psi(z,t) \, F_{\alpha\beta}F^{\alpha\beta} \, +  \, ...
\label{zeroaction}
\end{equation}
The function $\psi(z,t)$ is a localized function 
which is determined by the lump profile and
the bulk confinement; its precise form is not important for us. 
What is important is that $\psi$
evolves together with the lump.  In the limit when the lump vanishes, 
so does $\psi$.
This is obvious, since for the vanishing lump, there is no dual  
Meissner effect  that  protects the localized photon from the bulk 
confinement.  The linearized equation for the world-volume
zero mode photon on such a time-dependent background is
\begin{equation}
\nabla^2_{2+1} \, a_{\alpha} \, + \, {\dot{\kappa} \over \kappa} \, 
\dot{a}_{\alpha} \, = \, 0,
\label{eqa1}
\end{equation}
where $\nabla^2_{2 + 1}$ act only on the $2+1$-dimensional world-volume 
coordinates $x^{\alpha}$
and
\begin{equation} 
\kappa(t) \, = \, \int \, dz \, \psi(z,t).
\end{equation}
Taking a Fourier transform with respect to world-volume spatial coordinates, 
and keeping the explicit time dependence, we get
\begin{equation}
\ddot{a}_{\alpha}(k) \, + \, k^2 \, a_{\alpha}(k) \, + \, 
{\dot{\kappa} \over \kappa} \, \dot{a}_{\alpha}(k) \, = \, 0.
\label{eqa2}
\end{equation}
This is an equation for an oscillator with a time-dependent friction term. 
Note also that it coincides with the equation for a scalar field in a
$(2+1)$-dimensional  
Robertson-Walker universe, where the role of the scale factor is played by 
$\kappa(t)$ and $t$ plays the role of the conformal time coordinate.

Depending on the dynamics, we can distinguish three cases:
(1) $\kappa(t)$ decreases asymptotically approaching zero; 
(2) $\kappa(t)$ oscillates around zero;
(3) $\kappa(t)$ oscillates without crossing zero.
We shall consider these cases separately.  

{\bf (1) Asymptotically decreasing $\kappa$ }

In this case, generically there are
growing perturbations due to the negative friction term.  This growth
indicates instability with respect to particle production.  In
many cases the above equation can be integrated explicitly.  For
instance, for an exponentially decreasing $\kappa(t)$,
\beq
\kappa(t) \, = \, {\rm e}^{-\beta t},
\label{expkappa}
\eeq 
all the modes are exponentially growing as
\begin{equation}
a(k,t) \, \sim \, {\rm exp}\left({\beta \pm \sqrt{\beta^2 - 4k^2} \over 
2}t\right).
\label{expa}
\end{equation}

For a Gaussian time-dependence, $\kappa(t) \, = \, {\rm exp}(\beta^2t^2)$, 
Eq.~(\ref{eqa2}) is solved by Hermite polynomials, 
with $k^2 = 2n\beta^2$:
\begin{equation}
a(k,t) \, = \, H_n(t\beta) \, = \, (-1)^n \, {\rm e}^{(t\beta)^2} 
{d^n {\rm e}^{- (t\beta)^2} \over d(t\beta)^n},
\label{hermit}
\end{equation}
which grow as $H_n(t\beta) \sim t^n$.
However, other modes grow much faster.  For instance, 
the mode $k=0$ exhibits an "explosive" growth:
\begin{equation}
a_{\alpha}(0,t) \, = \,  \int _o^t \, {\rm e}^{(t'\beta)^2}\, dt'.
\label{explo}
\end{equation}
In fact, for the zero mode there is an explicit generic growing solution 
for an arbitrary decreasing $\kappa(t)$\footnote{This 
growth can also be understood as 
a parametric resonance effect by a suitable redefinition of the variables.  
We thank Lev Kofman for pointing this out to us and for valuable 
discussions on this issue.}:
\begin{equation}
a_{\alpha} (0, t) \, = \, \int_0^t  \,  {dt'  \over \kappa (t')}.
\label{explosion}
\end{equation}
This solution 
can be viewed as generation of a uniform electric field on the brane,
\begin{equation}
|E | \, = \, 1/\kappa(t).
\label{Ezeromode}
\end{equation}
After annihilation, this electric field can only survive in the form of 
straight,  
infinitely long strings.  At sufficiently late time, all modes in the
Gaussian model
grow like the zero mode 
(\ref{explo}).  This can be checked by verifying that 
(\ref{explosion}) gives an approximate solution to Eq.(\ref{eqa2}) 
for $t \, \gg \, k/\beta^2$.  The effect of these inhomogeneous modes is that
the electric field will vary from place to place, and the resulting strings 
will be rather irregular, resembling a cosmic string network formed in a
phase transition.

As our final example we consider a power-law dependence,
$\kappa(t) \, \sim \propto t^{-2\nu}$, with $\nu > 0$.  In this case, 
Eq.~(\ref{eqa2}) has a solution in terms of Bessel functions,
\begin{equation}
a(k,t) \, = \, t^{\nu + {1 \over 2}} Z_{\nu + {1 \over 2}}(kt),
\end{equation}
which grows like $t^{\nu}$ at late times.

{\bf (2) Oscillating $\kappa(t)$ and parametric resonance effects}

Such behavior would be expected in situations where the tachyonic
vacuum is at a finite distance from the origin.  If $\kappa$ goes
through zero during this oscillation, the friction term becomes
infinite and our approximation breaks down. Again, as in the case of
monotonically decreasing $\kappa$, there are growing mode solutions at
the begining of the evolution. For instance, the growing solution for the
zero mode (\ref{explosion}) is still valid, until $\kappa$ gets very small.
In the vicinity of $\kappa=0$, the time dependence should be well 
approximated by a linear function.  Choosing the origin of $t$ at the moment 
when $\kappa$ crosses zero, we then have 
\beq
\kappa(t)\approx Ct,
\label{smallt}
\eeq
and the solution of (\ref{eqa2}) is
\beq
a(k,t) \, = \, Z_0(kt).
\label{asmallt}
\end{equation}
Like the zero mode, these functions generally diverge at $t=0$. At some point
the energy in the gauge field becomes comparable to that of the soliton
itself, and back-reaction has to be taken into account.
Even before this happens, the time evolution of $\kappa(t)$ near $t=0$ 
may become too fast for the effective low-energy theory to apply. 

In some models, like the one discussed in the next section, $\kappa$
reaches zero, but instead of crossing over to negative values, it
starts growing again.  In such models,
Eqs.~(\ref{smallt}),(\ref{asmallt}) are replaced by
\beq
\kappa(t)\approx Ct^2,
\eeq
\beq
a(k,t) \, = \, t^{-1/2}Z_{-1/2}(kt),
\eeq
while the qualitative picture remains the same.

In the case when $\kappa$ oscillates
without ever crossing zero, the electric field (\ref{Ezeromode}) 
oscillates as well.
However, for inhomogeneous modes with $k\neq 0$ particle production can
occur due to
the parametric resonance effect.  This can be seen in
the following way. By making a substitution
\begin{equation}
a(k, t) \, = \, {u(k,t) \over \sqrt{\kappa (t)}}
\label{substitution}
\end{equation} 
we can bring equation (\ref{eqa2}) to the following form:
\begin{equation}
\ddot{u}(k,t) \, + \, \left [ \, k^2 - {1 \over 2} \left ( {\ddot{\kappa} 
\over \kappa} \, - \, {1\over 2} {\dot{\kappa}^2 \over \kappa^2} 
\right )\right ] \, u(k,t) \, = \, 0
\label{equ1}
\end{equation}
Since $\kappa(t)$ oscillates, so does the function in the square brackets. 
Thus, the equation
for $u(k,t)$ is an equation for a scalar field with a 
periodically changing mass.
It is well known that such systems exhibit parametric resonance 
behavior, the net result of which is that occupation numbers for
certain resonant 
values of $k^2$ grow exponentially\cite{kofmanlinde1}. Thus, an
explosive production of particles takes place.

\section{Explicit Model of Gauge Field Production}
  
We shall now make our analysis more concrete by considering a model in
which the gauge field localization on an unstable lump and its
subsequent intensive production can be demonstrated explicitly,
without need to consider the bulk confinement.  In this model, the
role of the bulk confinement is mimiced by a tree-level coupling of a
$U(1)$ gauge field to the lump profile.  This coupling renders the
gauge field outside the brane non-dynamical and produces a localized
photon.  We choose the action of the model in the form
\begin{equation}
\int \, d^3x dz \left [ {1 \over 2}\, (\partial_{\mu}\phi)^2  \, 
+ \, {\phi^2 \over 4}{\rm ln}(\phi^2) -
\phi^2 \, F_{\mu\nu}F^{\mu\nu} \, + \, ... \right ],
\label{actionphia}
\end{equation}
where $F_{\mu\nu}$ is the field strength of the $U(1)$ gauge field 
$a_{\mu}(x_{\alpha},z)$.  The crucial point is the tree-level coupling 
between $\phi$ and the gauge field. It is this coupling that
forces the  kinetic term of the gauge field to vanish outside the lump.  
Thus, from the low-energy perspective, such a  coupling to the lump 
has the same effect as the bulk confinement.  It makes the photon
a non-dynamical degree of freedom in the bulk.
  This similarity  is not surprising, since the above model
can be viewed as the effective low-energy theory obtained 
after integrating out the confining bulk
dynamics. Such an integration would in general generate other possible 
couplings between 
$\phi$ and $a$, including high-derivative interactions. 
We shall keep here the simplest coupling sufficient to produce a 
localized photon (note that a linear coupling in $\phi$ is forbidden by the 
discrete symmetry $\phi \, \rightarrow \, -\phi$).

We first consider the spectrum on a static lump background. 
The linearized equation for
the gauge field takes the form
\begin{equation}
\phi_{lump} \partial^{\nu}\, F_{\nu\mu} \,  -  \, 
2\acute{\phi}_{lump} F_{z\mu} \, = \, 0,
\label{eqa3}
\end{equation}
where primes denote $z$-derivatives. 
We perform the  standard separation of variables by demanding
\begin{equation}
 a_{\alpha}(z,x) \, = \, b^{(m)}(z) \, \tilde{a}_{\alpha}^{(m)}(x),~~~~~
 a_{z}(z,x) \, = \, \acute{b^{(m)}}(z) \, a_{z}^{(m)}(x)
\label{factorized}
\end{equation}
and introduce the new fields
\begin{equation}
  a_{\alpha}^{(m)}(x) \, = \, \tilde{a}_{\alpha}^{(m)}(x) \, - 
\, \partial_{\alpha} \, a_{z}^{(m)}(x),
\label{4dvector}
\end{equation}
which satisfy the $(2 + 1)$-dimensional massive vector field equation
\begin{equation}
 \partial^{\beta}F_{\beta\alpha}^{(m)}  \, = \,  -m^2 a_{\alpha}^{(m)}.
\label{eqb}
\end{equation}
Here, $F^{(m)}_{\beta\alpha}=\partial_\beta a^{(m)}_\alpha-\partial_\alpha
a^{(m)}_\beta$, indices $\mu,\nu$ take values from 0 to 3, corresponding to
the bulk spacetime, and indices $\alpha,\beta$ take values $0,1,2$ on
the worldsheet.  Due to eq.~(\ref{eqb}), all modes with $m \neq 0$ are 
transverse, $\partial^{\alpha} \, a_{\alpha}^{(m)}\, = \, 0$.

The functions $b^{(m)}(z)$ satisfy the following equation:
\begin{equation}
d_z^2 \, b^{(m)} \, - \,  z  \,  d_z \, b^{(m)} \, + \,  
m^2 \,  b^{(m)} \, =  \, 0. 
\label{eqa4}
\end{equation}
The solutions of this equation are Hermite polynomials
\begin{equation}
b^{(\sqrt{n})}(z) \, = \, H_n(z/\sqrt{2})
\label{hermit1}
\end{equation}
with $m \, = \, \sqrt{n}$, where $n$ is an integer.  
These functions are othonormalized, 
\begin{equation}
\int \, dz \,   {\rm e}^{-{z^2 \over 2} }\, H_n(z/\sqrt{2}) \, 
H_m(z/\sqrt{2}) \, = \, \sqrt{2\pi}  2^{n}\, n! \, \delta_{nm}
\label{normH}
\end{equation}
Thus, all the modes are localized on the lump. 
Notice that the mass levels grow as $m \sim \sqrt{n}$, as in the string 
spectrum,  as opposed to
$m \sim n$ that one would expect if the gauge fields were 
localized by compactification.
 
Now, to study the gauge field production during the
tachyon condensation, let us consider the linearized equation for the 
gauge modes on the time-dependent
background of Eq.(\ref{oscillator}),
\begin{equation}
\partial^{\alpha}\, F_{\alpha\mu} \,  - \partial^z \, F_{z\mu} \,  
+ \, 2 \, {\dot{A} \over A} \, F_{0\mu} \, + 
\, z F_{z\mu}
\,  = \, 0.
\label{eqa5}
\end{equation}
Since the $z$-dependence of the lump profile does not change in time in 
the first approximation,
we shall look for solutions in the factorized form given by 
(\ref{factorized}), with
$b^{(m)}(z)$ satisfying eq.(\ref{eqa4}).  For simplicity, we shall
look for solutions of (\ref{eqa5}) satisfying
the condition $a_0^{(m)} \, = \, \vec{\nabla} \, \vec{a}^{(m)} \, =
0$.  (Note that for $m=0$, this condition can be imposed due to the gauge
freedom, without loss of generality.)  
Performing a Fourier-transform with respect to the world-volume space momenta, 
we get the following equation for $a^{(m)}({\vec k},t)$
\begin{equation}
\ddot{a}^{(m)}_{\alpha}({\vec k},t) \, + \, (k^2 \, + \, m^2) \, 
a^{(m)}_{\alpha}({\vec k},t)  \, + \, 2 \, {\dot{A} \over A} \, 
\dot{a}^{(m)}_{\alpha}({\vec k},t) \, = \, 0.
\label{eqa6}
\end{equation}
This is nothing but equation (\ref{eqa2}), discussed in the previous section, 
with the substitutions $\kappa (t) \, \rightarrow \, A^2(t)$ and $k^2 \, 
\rightarrow \, k^2 \, + \, m^2$. Hence, all the results
of the previous section are applicable here. In particular, there is a
growing  solution for the ${\vec k}= m = 0$ mode,
\begin{equation}
\vec{a}^{(0)} (0,t) \, =    \, \vec{c} \,    \int_0^t \, {dt'  \over  
A (t')^2},
\label{explosion2}
\end{equation}
where $\vec{c}$ is an arbitrary constant vector along the world volume. 
This amounts to a growing uniform electric field. 
This solution breaks down when $A(t)$ approaches
zero and the electric field blows up. At that point the back reaction must 
be taken into account.

The behavior described above is obtained for the simplest possible 
coupling in the action
(\ref{actionphia}). With this choice, we have neglected all possible 
high-derivative interactions, such as
\begin{equation}
- \, \partial_{\gamma}\phi \, \partial^{\gamma}\phi \, 
F_{\mu\nu}F^{\mu\nu} \, .
\label{highd}
\end{equation}
Inclusion of such terms would replace the function $A^2(t)$ by some 
combination
of $A$ and its derivatives, which need not go through zero. 
In such a case, the equation for gauge quanta
may exhibit a parametric resonance behavior, as discussed at the end of 
the last section.
For example,  consider the effect of adding the term (\ref{highd}) 
to the original interaction
in (\ref{actionphia}). On the lump background, the gauge-kinetic term
now takes the form
\begin{equation}
- {\rm exp}(z^2/2) \, \left [ \, A^2 \, ( \, 1 \, - \,  (z/2)^2 \, ) \, 
+ \, \dot{A}^2 \right ] \, F_{\mu\nu}F^{\mu\nu}.
\label{ modifiedgaugek}
\end{equation}
Now the localizing function does not have a simple factorizable form, and 
the analysis of
$z$-dependent  excitations of the gauge field (the ones with $m^2 \, 
\neq \, 0$) becomes more
complicated.  For the zero mode,
the linearized equation takes the form (\ref{equ1}), 
where now $\kappa(t)$ is given by the
following equation:
\begin{equation}
\kappa(t) = \sqrt{{\pi \over 2}}\, \left [ \,  {3\over 4}  \, A^2 \, + \, 
\dot{A}^2 \, \right ].
\label{kappamod }
\end{equation}
This is an oscillatory function which never becomes zero.  Thus,  as 
discussed in the previous section, we expect the model to exhibit parametric 
resonance production of states with non-zero $k^2$.

\section{ Fate of the World-Volume States}

The question is what is the fate of the world-volume particles after
the lump completely annihilates?  The world-volume particles cannot
exist in the vacuum, since they are open string states.  In the
vacuum we can have either closed or infinitely long strings (electric flux
tubes).  Therefore, some part of the created open strings can be expected
to survive in this form.  
As we have seen, states of
different possible values of ${\vec k}$ are produced. Hence, we expect the
strings will be brownian, due to the amplification of short-wavelength
modes.  Not all of these will decay into closed strings.  The evolution
will be more like the evolution of cosmic strings \cite{book}, with the
characteristic length scale growing at the speed of light.

We would like to make a rough qualitative connection with the
production and evolution of cosmic string networks in ordinary
cosmology.  Consider a phase transition with a spontaneous breaking of
certain gauge $U(1)$ symmetry.  After the transition, the Universe becomes
superconducting, and the space gets populated by cosmic strings carrying
the $U(1)$ magnetic flux.  We distinguish two independent
sources of string production. First, any pre-existing magnetic
field (that was created before the Universe became superconducting)
gets trapped into the flux tubes.  In addition, magnetic flux
tubes are created by the Kibble mechanism, due to the winding of the Higgs
phase.  We shall ignore the second mechanism. Then, all the flux
tubes come from the pre-existing magnetic field, and the properties of the
string network will be determined by the mechanism that produced that
field.  The typical scale of the string network
will be given by the wavelength of the primodial magnetic field that gives
the highest contribution to the energy density. The above situation is 
analogous to our story.

The role of the phase transition with $U(1)$ breaking is played by the
brane annihilation. The difference is that the roles of magnetic and
electric fields are interchanged in our picture.  Instead of a
superconducting phase (Higgs phase), our Universe ends up in a
confining phase. As a result, we end up with electric flux tubes (or
fundamental strings in the $D$-brane picture) instead of magnetic
ones.  An important feature of brane annihilation is that it provides a
mechanism for efficient creation of the electric field.

In Appendix B we analyze the world-volume gauge field amplification in
the model with an exponential time dependence (\ref{expkappa}).  We
find that the energy density of the field grows as
\beq
\rho\sim (\beta/t)^{3/2}e^{\beta t},
\label{rho3}
\eeq
and the main contribution to the energy is given by the modes of momenta
\beq
k_*\sim (\beta/t)^{1/2}.
\label{k*}
\eeq
Because of this exponential growth, the gauge field energy becomes comparable
to that of the soliton itself on a timescale $\Delta t\sim ({\rm few}) 
\times \beta$.  At this point the back-reaction becomes important, and the 
field amplification terminates.  The characteristic length 
scale of the field variation at this time is $l_*\sim 1/k_* \sim
1/\beta$; we expect 
this to be the initial scale of the string network.

In this paper we have mostly discussed $(2+1)$-dimensional branes, but
the effect of gauge field amplification is quite generic, and in 
Appendix B we considered the case of $(d+1)$ dimensions with an
arbitrary $d$.  We found that in this general case Eq.~(\ref{rho3}) is
replaced by
\beq
\rho\sim \left({\beta\over{t}}\right)^{{d+1}\over{2}}e^{\beta t},
\eeq
while the characteristic momentum is still given by Eq.~(\ref{k*}).

At the beginning, the strings are located in the plane of the original
branes, but later they may depart from that plane as a result of
inter-string interactions.  We expect the evolution of strings to be similar
to that of a cosmic string network. Since the extra dimensions
are presumably compact, the strings will have gravitational effects
similar to the ordinary cosmic strings.  

 The string evolution may be modified in the presence of some additional
branes, which survive brane annihilation.  (Such branes must
be present in realistic scenarios: there should be at least
one surviving brane where the standard model particles are localized.)
What happens when a string hits one of the surviving branes?  This is
likely to result in some particle production, but the most dramatic
effect occurs if the gauge field trapped in the strings is massless 
on the brane.  In this case, the strings can end on the brane, and
a string segment passing through the brane will generally break into
two, with their ends attached to the brane.  Thus, the interaction of
strings with surviving branes can result in chopping up of the strings
into pieces ending on the branes.

Apart from producing a string network, the remaining part of the open
string energy will be annihilated in closed string states,
and in particular in gravitational radiation.  We have not studied
the back reaction of the created states on the tachyon condensation.  It
is certainly possible that this back reaction may resist the phase
transition and delay it, as in the case of parametric resonance
preheating \cite{kofmanlinde1}.

\section{$SU(5)$ Domain Walls as D-branes.}

 Let us now show that the $SU(5)$-domain walls discussed in 
Section 2 localize massless gauge fields in their world-volume by the
mechanism of ref. \cite{misha} discussed above.

 The key point is that part of the non-Abelian symmetry
$SU(3)_c \times SU(2)_L \times U(1)_Y$, that is unbroken away from the wall,
gets spontaneously broken in the wall. As we have seen, in the simplest wall
configuration, the Higgs field VEV vanishes in the core of the defect,
and the
$SU(5)$ symmetry is fully restored. In this solution, the only relevant
component of the adjoint Higgs field $\Phi$ is the one corresponding
to hypercharge direction (\ref{Y}), and all the other components are zero.
This, however, is not necessarily the energetically most favorable
configuration. In fact, other components
of the adjoint Higgs field $\Phi$, that vanish in the vacuum, may get
destabilized in the wall, triggering the symmetry breaking.
The fact that such destabilization can indeed happen was already shown
in \cite{tanmay}, and a more detailed analysis was performed in
\cite{tanmay1}. According to these studies, walls with broken symmetries in the
core have lower energy for all values of the parameters.
 For our purposes, it is enough to show that such a situation occurs at
least for some values of the parameters. 

 To see that symmetry breaking inside the wall is possible,
consider the Higgs potential with $\Phi$ restricted to lie
along the diagonal
$$
\Phi = a \lambda_3 + b \lambda_8 + c \tau_3 + v Y,
$$
where $\lambda_3$ and $\lambda_8$ are matrices from the $SU(3)_c$ Cartan
subalgebra, $\tau_3$ is the weak isospin and $Y$ is the hypercharge 
generator:

\begin{equation}
\lambda_3=\frac{1}{2} {\rm diag}(1,-1,0,0,0) \ ,  
\label{lambda3}
\end{equation}
\begin{equation}
\lambda_8=\frac{1}{2\sqrt{3}} {\rm diag}(1,1,-2,0,0) \ ,
\label{lambda8}
\end{equation} 
\begin{equation}
\tau_3=\frac{1}{2} {\rm diag}(0,0,0,1,-1) \ ,  
\label{tau3}
\end{equation}
\begin{equation}
Y=\frac{1}{2\sqrt{15}} {\rm diag}(2,2,2,-3,-3) \ ,
\label{ymatrix}
\end{equation}
 all matrices being normalized to unity.
For the simplest wall, only the $Y$-component is non-zero.
Let us now show that, at least for a region of the parameter space,
 some other component(s) among $\lambda_3$,$\lambda_8$ and $\tau_3$ 
(which vanish in the vacuum) can pick up a VEV and break the 
gauge symmetry inside the wall in a different fashion.  This can be 
simply understood by examining the linearized Schrodinger
equation for small excitations, $\epsilon = \epsilon_0 (x) e^{-i\omega t}$
($\epsilon$ is either the $a,b$ or $c$ component), in the wall background
\begin{equation}
\left [-\partial_x^2 + (-m^2 + (v(x))^2 (h + \lambda r)) \right]\epsilon_0
= \omega^2 \epsilon_0,
\label{schrod}
\end{equation}
where the wall is taken to lie in the $x=0$ plane, and
for $\epsilon$ being in the $a,b,c$ directions we have 
$r = 2/5, ~ 2/5 ~ {\rm and} ~ 9/10$, respectively. 
The function $v(x)$ is the profile of the wall, which for a planar infinite 
wall can be approximated by the kink solution,
\begin{equation}
v(x) = {m \over {\sqrt{\lambda '}} } {\rm tanh}
               \left ({{mx} \over {\sqrt{2}}} \right ).
\label{profile}
\end{equation}
 It is obvious that there is a wide region of the parameter space for which
the above Schrodinger equation has bound state (negative $\omega^2$) solutions.
For instance, it is enough to take
\begin{equation}
 (h + \lambda r) = \lambda '
\label{condition}
\end{equation}
 In such a situation, the symmetry inside the wall will be less than 
the symmetry
outside. To be concrete, we shall take an example of a stable wall 
found in \cite{tanmay1}, in which the Higgs VEV inside is
diag$(1,-1,0,1,-1)$ and thus the symmetry is $SU(2)\otimes SU(2)
\otimes U(1)_{color}\otimes U(1)_{weak}$. Here,
the subscipts indicate that the corresponding
$U(1)$-s belong to the $SU(3)$ and $SU(2)$ subgroups of the symmetry
group that is unbroken in the vacuum outside the brane.
 Let us concentrate on one of them, say $U(1)_{color}$.
The corresponding photon gauge field is a $\lambda_8$-gluon $g^8_{\mu}$
of QCD, and becomes
a part of the confining $SU(3)$-theory outside the brane. But on the
brane the $SU(3)$ is in the Higgs phase and the $\lambda_8$-gluon is a part of
an Abelian symmetry in the coulomb phase. Thus, the $SU(5)$ domain wall
supports a massless gluon excitation in its world volume.

 We can consider a wall in which there is a uniform  
$g^8_{\mu}$-electric field. Such a wall can anihhilate with an anti-wall
of opposite topological charge that carries no electric field.
The result of the anihhilation will be a set of infinitelly long electric flux
tubes localized in the plane of the original wall. In the absence of
fundamental charges, the flux tubes are
stable due to charge conservation.  This is closely analogous 
to what happens in anihhilation
of a $D(p)$-brane system with a Born-Infeld electric field.

\section{Appendix A}

 In this Appendix we shall briefly review the central extension 
of $N=1$ globally supersymmetric algebra and its implications for
the domain wall solution. We shall mostly follow \cite{misha}.
In $N=1$ SUSY, localized objects (e.g.
particles or monopoles) cannot be BPS saturated, but domain walls
can. The reason is that in the wall background the 
$4$D $N=1$ SUSY algebra admits
the central extension,

\begin{equation}
\bigg\{\bar{Q}_{\alpha}Q_{\beta} \bigg\} = 2(\gamma^{\mu}P_{\mu})
_{\alpha\beta}+ 2\left(\gamma^5\sigma_{\mu\nu}J^{\mu\nu}\right)_{\alpha\beta},
\label{algebra}
\end{equation}
where $\sigma_{\mu\nu}={1\over 2} [\gamma_\mu\gamma_\nu]$ and
\begin{equation}
J^{\mu\nu} = \int d^3x \epsilon^{0\mu\nu\omega}\partial_{\omega}W
\end{equation}
is the trivially conserved topological charge. Here, $W$ is the superpotential.
Due to this central extension, backgrounds with non-zero $J^{\mu\nu}$
may preserve half of the original supersymmetry (and thus be
BPS-saturates states), even though the energy
is non-zero. Such configurations are domain walls across which
superpotential changes. Let us consider theories with a single chiral
superfield $\Phi$. We shall denote the scalar component of the superfield
by the same symbol, and the fermionic member by $\psi$. We choose the basis
in which all the $\gamma_{\mu}$ matrices are real.
 For static field configurations that depend on a single
space coordinate $z$, the energy functional has the form
\begin{equation}
\int dz \left ( K_{\Phi^*\Phi} \partial_z\Phi^* \partial_z\Phi^*
+  K_{\Phi^*\Phi}^{-1} W_{\Phi^*}^* W_{\Phi} \right ),
\end{equation}
where $K(\Phi\Phi^*)$ is the K\"ahler function, and subscripts denote
derivatives with respect to the fields. Rewriting the above as a perfect
square and integrating the rest, we get
\begin{equation}
\int dz  K_{\Phi^*\Phi} \left | \partial_z\Phi
\pm  K_{\Phi^*\Phi}^{-1} W_{\Phi^*}^* \right |^2  \mp
\left (W(+\infty) - W(-\infty)\right ).
\label{energyBPS}
\end{equation}
For the BPS-saturated wall the first term
vanishes,
\begin{equation}
\partial_z\Phi \pm  K_{\Phi^*\Phi}^{-1} W_{\Phi^*}^* = 0,
\label{BPScondition}
\end{equation}
and the tension is given by
\begin{equation}
T = W(+\infty) - W(-\infty).
\label{tensionBPS}
\end{equation}

The important result is that, due to non-renormalization of the
superpotential, this tension is exact to all orders in perturbation theory.
The profile of the wall can receive radiative corrections due to
renormalization of K\"ahler, but the tension is fixed once and for all.
The BPS condition (\ref{BPScondition}) 
guarantees that half of the supersymmetric
transformations annihilate the wall. As a result, the low-energy
world volume theory
is a $(2+1)$-dimensional $N=1$ supersymmetric theory. In the case of a single
chiral superfield, it consist of a real zero mode scalar and a
fermion. $z$-dependence of their profiles must be identical, due to unbroken
supersymmetry. This can be seen explicitly for an arbitrary
$W$ without actually solving the equations. Let us for simplicity restrict
the analysis to 
real solutions. Then the wall is given by a profile $\Phi_w(z)$, which
satisfies equation (\ref{BPScondition}). The profile of the
scalar zero mode is simply given by $\Phi_w(z)'$, and thus by
$W_{\Phi}(\Phi_w)$. The profile of the fermionic zero
mode satisfies the equation
\begin{equation}
(\gamma_z\partial_z + W_{\Phi\Phi}(\Phi_w))\psi(z) = 0, 
\end{equation}
which is solved by 
\begin{equation}
\psi(z) = \epsilon exp(\pm\int_0^z  W_{\Phi\Phi}(z')dz'),
\label{fermion}
\end{equation}
where $\epsilon$ is an eigenspinor of $\gamma_z\epsilon = \mp\epsilon$.
Differentiating the BPS condition (\ref{BPScondition}) and integrating
with respect to $\Phi$, we get
\begin{equation}
\int_0^z  W_{\Phi\Phi}(z')dz' = {\rm ln}
\left({\Phi_w(z)'\over\Phi_w(0)'}\right).
\end{equation}
Plugging this back into (\ref{fermion}) we obtain
\begin{equation}
\psi(z) = {\Phi_w(z)'\over\Phi_w(0)'}.
\label{fermibose}
\end{equation}
Thus, the wave functions of (canonically normalized) fermionic and bosonic
zero modes are identical, as they should be by supersymmetry.
\vspace{0.5cm}

\section{Appendix B}

In this Appendix we shall study the world-sheet field amplification
during the unstable brane decay.  For simplicity, we shall consider a
masless scalar field.  As we saw in Section 5, it is described by the
same field equation as the gauge field; thus, the two cases are
equivalent, apart from the number of components.  The corresponding 
effective action is
\beq
S = \int dt~ d^d x~ \kappa(t)~ \partial_\mu\varphi\partial^\mu\varphi,
\label{effective}
\eeq
where $d$ is the number of spatial dimensions of the worldsheet.

We shall adopt a simple model where $\kappa (t) =1$ at $t<0$ and
\beq
\kappa (t)=e^{-\beta t}
\eeq
at $t>0$.  For $t<0$, $\varphi$ satisfies the usual free scalar field 
equation, and we can represent the field operator as
\beq
\vp (x)=\sum_k \left(a_k\vp_k(t)e^{i{\bf kx}} +a_k^{+} \vp_k^*(t) 
e^{-i{\bf kx}}\right),
\label{phiop}
\eeq
with
\beq
\vp_k(t)={1\over{\sqrt{2kV}}}e^{-ikt},
\eeq
where $V$ is the $d$-dimensional normalization volume, to be taken to
infinity at the end.

At $t>0$, the mode functions are linear combinations of the solutions
(\ref{expa}), 
\beq
\vp_k(t)=A_k\exp (\np t)+B_k\exp (\nm t),
\label{phik}
\eeq
where 
\beq
\nu_k^{(\pm)}={1\over{2}}(\beta\pm\sqrt{\beta^2-4k^2}).
\eeq
The coefficients $A_k$ and $B_k$ can be found by matching the mode
functions and their derivatives at $t=0$.  This gives
\beq
A_k=-{\nm+ik\over{\sqrt{2kV}(\np-\nm)}}
\label{Ak}
\eeq
and
\beq
B_k={\np+ik\over{\sqrt{2kV}(\np-\nm)}}.
\label{Bk}
\eeq

The expectation value of the energy density in the in-vacuum state is
given by
\beq
\rho={\kappa(t)\over{2}}\sum_k ({\dot\vp}_k {\dot\vp}_k^* +k^2\vp_k \vp_k^*).
\label{exprho}
\eeq
The effective field theory (\ref{effective}) is valid only up to some
scale $k\sim M$, so the summation in (\ref{exprho}) has to be cut off
at that scale.  We should also subtract the zero-point energy of the
field prior to the decay,
\beq
\rho_0={1\over{2}}\sum_k k,
\eeq
with the same cutoff.  This is not important, however, since we are
interested only
in the time-dependent part of $\rho$.  

It is clear from the mode function solutions (\ref{phik}) that the
fastest growing modes are those with the lowest $k$.  Hence, we expect
the low-$k$ modes to give the dominant contribution at late times.
For $k<\beta/2$, both $\np$ and $\nm$ are real, and we obtain the following
expression after substituting
(\ref{phik})-(\ref{Bk}) into (\ref{exprho}):
\beq
\rho={1\over{2}}e^{-\beta t}\sum_k{1\over{2kV(\beta^2-4k^2)}}
[(\np)^2+k^2] [(\nm)^2+k^2]\left(e^{2 \np t}+e^{2\nm t}\right).
\label{rhot}
\eeq
As expected, al large $t$ this sum is dominated by small $k$.
Expanding $\nu_k^{(\pm)}$ in powers of $k$, keeping only the
leading terms, and replacing summation by integration, we have
\beq
\rho\approx e^{\beta t}\int {d^d
k\over{(2\pi)^d}}ke^{-(4k^2/\beta)t}
\sim \left({\beta\over{t}}\right)^{{d+1}\over{2}}e^{\beta t}.
\label{rho2}
\eeq
The dominant contribution to the integral is given by the modes with 
\beq
k\sim (\beta/t)^{1/2}.
\eeq

{\bf Acknowledgments}
\vspace{0.1cm} \\

We would like to thank C. Deffayet, G. Gabadadze, J. Garriga, L. Kofman, and
M. Kolanovic for useful discussions on the subject of the
paper. Special thanks are to A. Gruzinov for valuable conversations on the
dynamics of various non-linear systems.  The work of G.D. is supported
in part by a David and Lucile Packard Foundation Fellowship for
Science and Engineering, by Alfred P. Sloan Foundation fellowship, and
by NSF grant PHY-0070787.  The work of A.V. is supported in part by
the NSF.  G.D.  acknowledges the hospitality of ICTP
(Trieste, Italy), and of the 
Aspen Center for Physics, where part of this
work was done.

\end{document}